# SPATIAL VARIATION OF TOTAL COLUMN OZONE ON A GLOBAL SCALE

By Michael L. Stein[1]

*University of Chicago*

The spatial dependence of total column ozone varies strongly with latitude, so that homogeneous models (invariant to all rotations) are clearly unsuitable. However, an assumption of axial symmetry, which means that the process model is invariant to rotations about the Earth's axis, is much more plausible and considerably simplifies the modeling. Using TOMS (Total Ozone Mapping Spectrometer) measurements of total column ozone over a six-day period, this work investigates the modeling of axially symmetric processes on the sphere using expansions in spherical harmonics. It turns out that one can capture many of the large scale features of the spatial covariance structure using a relatively small number of terms in such an expansion, but the resulting fitted model provides a horrible fit to the data when evaluated via its likelihood because of its inability to describe accurately the process's local behavior. Thus, there remains the challenge of developing computationally tractable models that capture both the large and small scale structure of these data.

**1. Introduction.** Random process models on the sphere go back at least to Obukhov (1947) [20] (see [32], Vol. 2, page 133), who derived the general spectral representation for a (weakly) homogeneous process on the sphere in terms of spherical harmonics. By homogeneous (sometimes referred to as isotropic for processes on the sphere), we mean that the first two moments of the process are invariant under all rotations of the sphere. Homogeneous models have found extensive applications in models for the geomagnetic potential [18, 19]. For atmospheric processes, invariance to all rotations will

Received January 2007; revised March 2007.

[1] Although the research described in this article has been funded wholly or in part by the United States Environmental Protection Agency through STAR cooperative agreement R-82940201-0 to the University of Chicago, it has not been subjected to the Agency's required peer and policy review and therefore does not necessarily reflect the views of the Agency and no official endorsement should be inferred.

generally be too strong an assumption because of differences in how the process behaves at different latitudes. However, invariance to rotations about the Earth's axis may sometimes hold to a decent approximation, especially for processes in the stratosphere for which surface effects may not matter much. Jones [13] called such processes axially symmetric and showed how they could be represented in terms of spherical harmonics. This paper applies Jones' approach to satellite-based observations of total column ozone over a six-day period. The near global coverage and high spatial resolution of the available satellite data provides a good testbed for examining the effectiveness of this approach.

Section 2 describes the Level 2 (ungridded) TOMS ozone data used here and briefly reviews past efforts to model statistically total column ozone on a global scale. Section 3 gives some preliminary analyses, including the approach to removing a mean function from the observations and various displays of empirical variograms for the ozone residuals. These analyses show that there are major differences in the spatial dependence as a function of latitude, so that homogeneity is badly untrue, but that axial symmetry is fairly reasonable and, thus, an axially symmetric model appears to provide a good compromise between fidelity and complexity. Section 4 explains how series expansions in spherical harmonics can be used to represent axially symmetric processes in terms of the covariance matrices of the random coefficients. Truncating this series expansion after a moderate number of terms leads to major computational advantages in terms of calculating kriging predictors and likelihoods. Section 5 estimates the parameters in such a truncated expansion for six days of TOMS data. In order to obtain an estimated covariance structure that follows the large scale patterns in the empirical variogram, a weighted least squares fit is found. The resulting fitted model provides good visual agreement with the empirical variograms as a function of latitude. Unfortunately, as demonstrated in Section 6, the weighted least squares fit has a horribly low likelihood even compared to a model that assumes no spatial dependence, apparently because of its inability to capture the local behavior of the data. Section 7 discusses how one might obtain a better compromise between computational feasibility and fidelity to the data. In addition, it describes a new data product that NASA might consider producing that would be intermediate between the Level 2 data used here and the Level 3 daily gridded data product it now produces. This product would give values of ozone on a grid but retain the time of observation information in the Level 2 data, giving multiple observations in a day for parts of the Earth scanned more than once during the day. There is a considerable literature on mapping surface levels of biologically harmful ultraviolet-B radiation on time scales as short as 15 minutes [1, 29], for which total column ozone levels are a critical input, so an understanding of variations in total column ozone on time scales shorter than a day is of obvious interest.

**2. Data.** The Nimbus-7 satellite carried a TOMS instrument that measured total column ozone daily from November 1, 1978 to May 6, 1993. This satellite followed a Sun-synchronous polar orbit with an orbital frequency of 13.825 orbits a day (about 104 minutes). As the satellite orbited, a scanning mirror repeatedly scanned across a track about 3000 km wide, each track yielding 35 total column ozone measurements [16]. This version of the data is known as Level 2 and is publicly available from http://disc.sci.gsfc.nasa.gov/data/datapool/TOMS/Level_2/. This work focuses on data from six consecutive days, May 1–6, 1990, containing over one million observations. Because the instrument makes use of backscattered sunlight, measurements are not available south of 73°S during this week.

It is in fact more common to study the gridded Level 3 version of the TOMS data, in which the Level 2 values are interpolated daily onto a grid of 1° latitude × 1.25° longitude for latitudes between 50°S and 50°N (with fewer grid points per latitude near the poles) [16]. For example, [5, 8, 12] and [26] consider statistical models for the spatial–temporal variation of various subsets of the Level 3 data. To a rough approximation, Level 3 observations can be treated as if they were taken at local noon [5]. However, the specific times of observations are lost in the Level 3 data. Thus, with Level 3 data, it is not possible to distinguish small-scale spatial (on the order of hundreds of kilometers) from small-scale temporal (on the order of a few hours) variation. In contrast, Level 2 data does provide some information for distinguishing between these sources of variation. In particular, for regions not near the equator, there is considerable overlap between the observation domains for consecutive orbits, so there are many pairs of measurements at nearly the same spatial location that are about 104 minutes apart. The focus in this work is on the purely spatial variation at a fixed time. We will study this variation by only considering dependence between observations on a common orbit, since observations within an orbit, especially those that are geographically close, were taken within minutes of each other and are effectively simultaneous.

Cressie and Johanneson [4] is the only other work of which I am aware that considers statistical models for Level 2 TOMS data on a global scale. In addition, as we do here, they use a series expansion for the spatial covariance function. However, they do not distinguish between observations on different orbits within the same day and they do not attempt to take advantage of axial symmetry in their model.

**3. Preliminary analyses.** Figure 1 shows boxplots of all observed total column ozone levels by latitude bands for the period May 1–6, 1990. This figure shows that ozone levels vary strongly with latitude and that variation within a latitude is much lower near the equator than elsewhere. There is also some evidence of positive skewness even on the logarithmic scale for the distributions of ozone levels between the equator and 40°N. As a general principle in analyzing space–time data, I would argue that it is important to separate out purely spatial variation (i.e., not varying with time) from temporally varying spatial patterns (see [28] for an example of problems that can happen if this is not done). Here, we will use a regression approach to remove at least some of this purely spatial variation. Obviously, the results then depend somewhat on the regressors chosen, but a somewhat arbitrarily chosen mean function model seems preferable to making no attempt to remove the purely spatial variation. This and all further analyzes here will be done on the natural logarithms of total column ozone (in Dobson units), since the process is more nearly Gaussian on this scale.

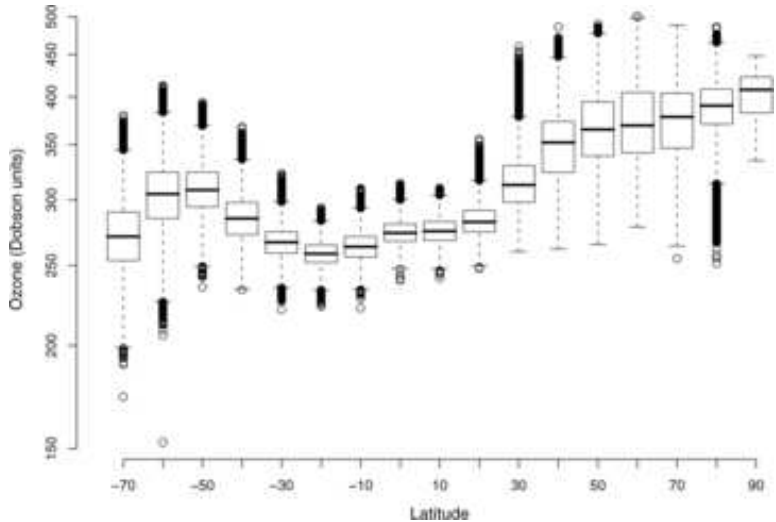

Fig. 1. *Total column ozone by latitude band, May 1–6, 1990.*

The regressors used were selected spherical harmonics. Specifically, denoting latitude by $L$ ($-\frac{1}{2}\pi \leq L \leq \frac{1}{2}\pi$) and longitude by $\ell$ ($-\pi < \ell \leq \pi$) and writing $P_n^m$ for the Legendre polynomial of degree $n$ and order $m$, the regressors were $P_n^m(\sin L)\cos(m\ell)$ and $P_n^m(\sin L)\sin(m\ell)$ for $n = 0, 1, \ldots, 12$ and $m = 0, \ldots, \min(3, n)$ for a total of 78 covariates. The restriction $m \leq n$ just reflects the fact that $P_n^m$ is identically 0 for $m > n$ and the restriction $m \leq 3$ was imposed to allow more flexibility in variations across latitudes than across longitudes within latitudes. The observations were averaged into bins of size 1° latitude by 2° longitude and then the coefficients of the mean function were estimated by ordinary least squares using the bin averages of ozone as the responses and the spherical harmonics evaluated at the bin averages of latitudes and longitudes as the regressors. The fitted model explains 88% of the variation in the original ozone observations (even though the model was fitted to the bin averages). If one uses a similar approach using only covariates that do not depend on longitude, the fraction of variation explained is about 80%, so including some dependence on longitude in the mean function seems justified. A boxplot of the residuals by latitude from the model including longitudinal dependence (not shown) indicates that the latitudinal trend has been effectively removed and the skewness much reduced, but these residuals still show substantially more variation as one moves poleward. Separate plots for the residuals on each of the six days are all quite similar, indicating that the greater variation toward the poles is a stable pattern during this time period.

The spatial variation of these residuals clearly depends on latitude, which any model for the spatial covariance structure will need to capture. However, one might hope that the spatial dependence of residual ozone at two sites depends on the longitudes of these sites only through their difference, an invariance that, following [13], we will call axial symmetry. Specifically, consider a random field $Z$ on the sphere with coordinates designated by $(L, \ell)$. We will call $Z$ (weakly) axially symmetric if its mean depends only on latitude and there exists a function $K$ on $[-\frac{1}{2}\pi, \frac{1}{2}\pi]^2 \times (-\pi, \pi]$ such that, for all $(L, \ell)$ and $(L', \ell')$,

$$\text{(1)} \qquad \text{cov}\{Z(L, \ell), Z(L', \ell')\} = K(L, L', \ell - \ell').$$

We will call such a $K$ an axially symmetric covariance function.

For examining the local variation of a spatial process, it is often useful to consider the variogram rather than the covariance function. For an axially symmetric process, there exists a function $\gamma$ on $[-\frac{1}{2}\pi, \frac{1}{2}\pi]^2 \times (-\pi, \pi]$ such that, for all $(L, \ell)$ and $(L', \ell')$,

$$\text{(2)} \qquad \tfrac{1}{2}\text{var}\{Z(L, \ell) - Z(L', \ell')\} = \gamma(L, L', \ell - \ell').$$

Since we have only required that the mean of $Z$ be independent of longitude, we do not necessarily have $E\{Z(L, \ell) - Z(L', \ell')\}^2 = \text{var}\{Z(L, \ell) - Z(L', \ell')\}$ as we do for stationary processes. It is not clear whether one should define the variogram to be $\frac{1}{2}E\{Z(L, \ell) - Z(L', \ell')\}^2$ or $\frac{1}{2}\text{var}\{Z(L, \ell) - Z(L', \ell')\}$ in this case, but since we work with residuals here, the difference should not matter much. Perhaps a more important issue is that, even if $\text{var}\{Z(L, \ell)\} = K(L, L, 0)$ is uniformly bounded in $L$, in contrast to the situation for stationary processes on $\mathbb{R}^d$ (or homogeneous processes on the sphere), it does not follow that one can identify $K$ up to an additive constant from $\gamma$. Specifically, it is possible to show that two axially symmetric covariance functions $K$ and $K_1$ yield the same variogram if and only if their difference, $K(L, L', \ell) - K_1(L, L', \ell)$, can be written in the form $a(L) + a(L')$ for some function $a$. Consequently, one cannot compute ordinary kriging predictors (best linear unbiased predictors in which the mean is assumed to be an unknown constant) just from $\gamma$. Despite this difficulty, the fact that $\gamma(L, L, 0) = 0$ for all $L$, whereas $K(L, L, 0)$ may vary with $L$, makes it easier to visualize differences in local variation as a function of latitude using the variogram.

Figure 2 gives contour plots of an empirical version of $\gamma(L, L', \ell)$ for $L = 40°$S, $0°$N, $20°$N, $40°$N and $60°$N, for $|L - L'| < 9°$ and $|\ell| < 20°$. These empirical variograms use only pairs of observations from the same orbit, so these plots should be nearly unaffected by temporal variations in $Z$. The Appendix gives further details on how the empirical variograms were computed. The plots show dramatic variations in $\gamma$ as the latitude of the first observation, $L$, varies. At longer lags, there is generally much more variation for more poleward latitudes. Furthermore, the patterns in the Northern and Southern hemispheres are quite different, as can be seen by comparing the results at $40°$N and $40°$S. Another important feature of these figures is the local anisotropies. For example, at $40°$S, the variogram increases much more quickly as $L'$ moves northward rather than southward. The approximately elliptical contours at $40°$N demonstrate a kind of local geometric anisotropy, indicating greater dependence in the southwest–northeast direction than in the northwest–southeast direction. I will say a process is longitudinally reversible if $K(L, L', \ell) = K(L, L', -\ell)$ for all $L, L'$ and $\ell$; it is apparent that the residual ozone process does not possess this property.

For these plots to represent a sensible summary of the spatial variation, $\text{var}\{Z(L, \ell, t) - Z(L', \ell', t)\}$ should be, at least approximately, independent of $t$ over this six-day period and depend on the longitudes $\ell$ and $\ell'$ only through their difference. Figure 3 gives pairs of spatial variograms along the east–west direction at the same latitudes as in Figure 2, one using orbits in which the first observation for the orbit was in the Western hemisphere and the other using orbits with first observation in the Eastern hemisphere. The two variograms within a latitude band do differ somewhat, although they are generally much more similar to each other than they are to the variograms at different latitudes. Other variogram plots (not shown) exhibit a similar consistency in results for the Western and Eastern hemispheres and across the six days, suggesting that axial symmetry is a good if not perfect assumption for these data.

One could get further information about spatial variation, especially about variation at larger differences in longitude, by considering pairs of observations from different orbits. However, the observations would then not be so close together in time. To investigate whether this time difference matters, for integer $t$, define $\hat{\gamma}(L, L', \ell, t)$ as follows: set $\hat{\gamma}(L, L', \ell, 0) = \hat{\gamma}(L, L', \ell)$ as in the Appendix and, for $t \neq 0$, define $\hat{\gamma}(L, L', \ell, t)$ similarly, except that the average squared difference is taken over pairs of observations in which the second observation is $t$ orbits away from the first observation. Figure 4 plots $\hat{\gamma}(L, L, \ell, t)$ for $t = -1, 0, 1$ and $L = 0°$N and $60°$N. At $60°$N, for all but the shortest positive longitudinal lags $\ell$, $\hat{\gamma}(L,L,\ell,1) < \hat{\gamma}(L,L,\ell,0)$ and the inequality is reversed for $\ell < 0$, which one would expect if ozone tends to move eastward. At 0°N, because orbits just barely overlap, we only get to see "half" of the picture. Nevertheless, for the shorter lags, it now appears that the pattern has been reversed, indicative of westward movement of ozone. Both the eastward movement at 60°N and the westward movement at the equator were also found in [5] by looking at Level 3 TOMS data on a daily basis. It is interesting to find that these effects can be seen here looking at observations on consecutive orbits, which are only 104 minutes apart. It is not possible to look for such effects on this shorter time scale using the Level 3 data.

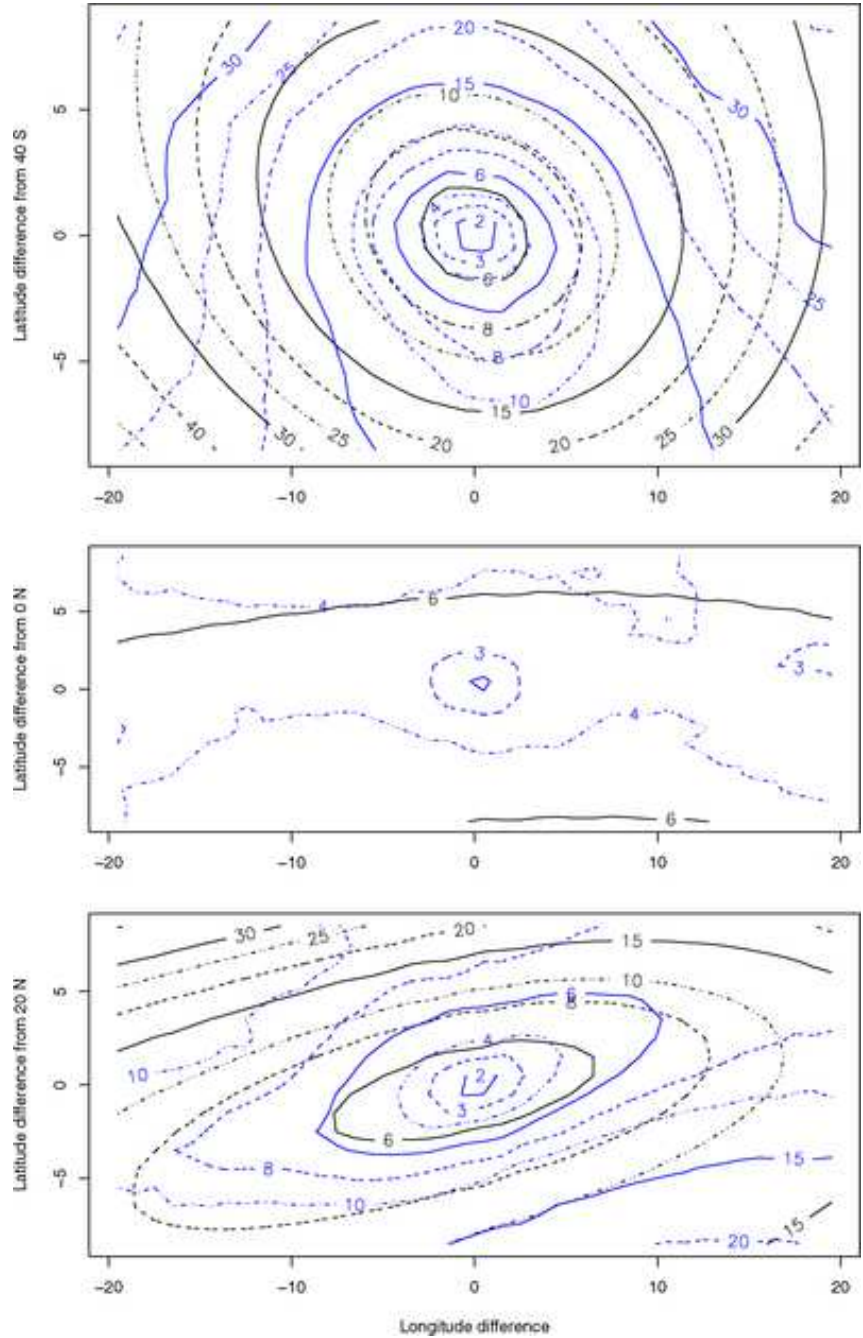

Fig. 2. *Empirical (blue) and fitted (black) variograms of residual ozone with first observation at, respectively, $40°S$, $0°N$, $20°N$, $40°N$ and $60°N$. Actual values are $10^{-4}$ times the displayed contour levels. The aspect ratios of the plots vary with latitude so that they roughly correspond to the local relationship between a degree latitude and a degree longitude.*

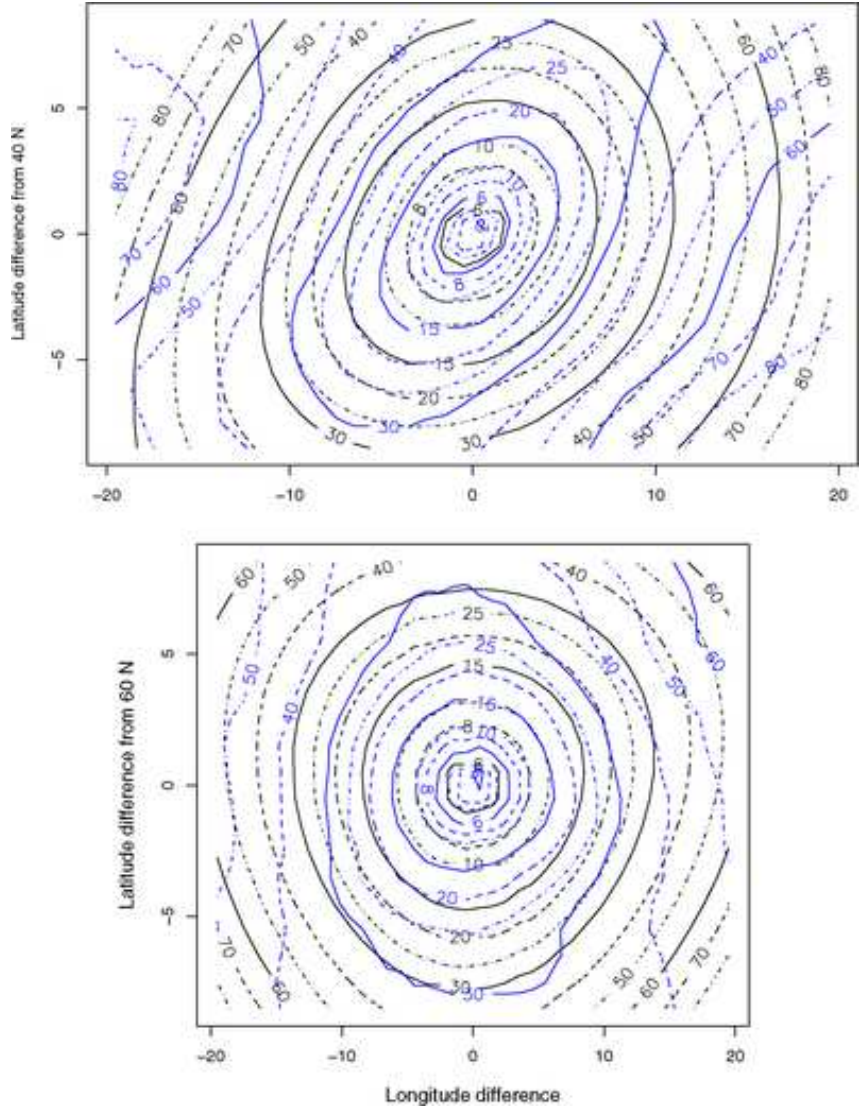

Fig. 2. *(Continued.)*

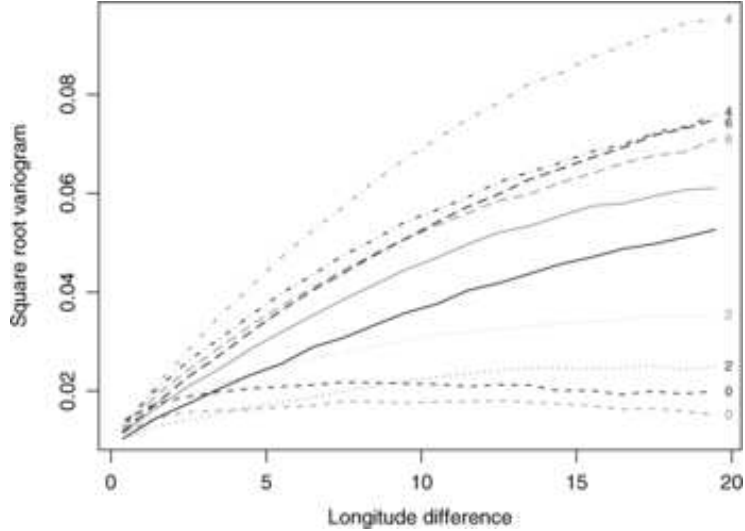

Fig. 3. *Empirical variograms across longitudes for various latitudes, $40°S$ (unlabeled), $0°N$, $20°N$, $40°N$ and $60°N$ (labeled 0, 2, 4 and 6, respectively) for orbits with first observation in the Eastern hemisphere (black) and Western hemisphere (gray).*

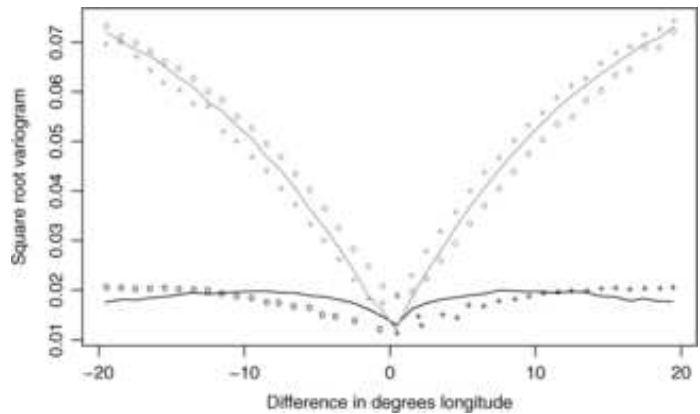

Fig. 4. *Variograms within latitude band as a function of difference in longitudes $(\ell_1 - \ell_2)$ at $0°N$ (black) and $60°N$ (gray) for pairs of observations in the same orbit (solid curves), second observation in orbit before first ($\circ$) and second observation in orbit after first ($+$).*

**4. Model.** There are a number of ways one might seek to model axially symmetric processes. One possibility would be to adapt an approach for modeling nonstationary covariance functions and try to restrict it so that only axially symmetric processes result. Sampson and Guttorp [23] describe a popular method for modeling and estimating the nonstationary covariance function of a process on the plane by assuming the process is isotropic after some mapping of the plane to itself. One could try to adapt this approach to obtaining axially symmetric models by starting out with a homogeneous process on the sphere and then allowing deformations that retain axial symmetry; for example, for homogeneous $Z$ on the sphere, consider $Z(\phi(L), \ell + \theta(L))$ for $\phi$ a smooth, increasing function from $[-\frac{1}{2}\pi, \frac{1}{2}\pi]$ to itself and $\theta$ continuous. However, this approach does not appear to allow for the strong variation across latitudes in $\gamma(L, L, \ell)$ (i.e., the variation across longitudes within a latitude) shown in Figure 2. Paciorek and Schervish [21] describe a method of generating nonstationary spatial covariance functions that can be adapted to produce axially symmetric covariance functions (e.g., by letting $\Sigma_i$ in (5) of [21] depend only on latitude). The approach of Jun and Stein [14] is specifically aimed at producing axially symmetric (space–time) covariance functions for processes on spheres. However, unlike the series approach used here, the approaches in [21] and [14] can only produce some subset of the class of axially symmetric models. In particular, it is not clear they could capture the radically different variogram structure at longer spatial lags as latitude varies shown in Figure 2 at the same time as preserving the fairly similar local (up to a few hundred kilometers) variogram structure at different latitudes.

This work adopts the approach described by Jones [13] (Section 4) using expansions in terms of spherical harmonics. Jones [13] writes everything in terms of real quantities, but, as is often the case, it is cleaner to use a complex representation, even when one is only interested in real processes, and we shall do so here. Write $\bar{P}_n^m$ for the normalized version of $P_n^m$ (normalized so its squared integral on $[-1, 1]$ is 1). Consider

$$(3) \qquad Z(L, \ell) = \sum_{n=0}^{\infty} \sum_{m=-n}^{n} Y_{nm} e^{im\ell} \bar{P}_n^m(\sin L),$$

where the $Y_{nm}$'s are complex-valued random variables and the infinite sum is understood to converge in mean square. The integer $m$ gives the longitudinal frequency and is generally called the wavenumber in the geophysical literature. If $Y_{nm} = Y_{n,-m}^*$ (where $^*$ indicates complex conjugate), then $Z$ is real-valued and we shall assume this holds hereafter.

One reason for using (3) is that the second-order structure of the $Y_{nm}$'s is particularly simple if $Z$ is homogeneous on the sphere. Define $\delta_{ab}$ to be 1 if $a = b$ and 0 otherwise. For $Z$ to be a real mean square continuous homogeneous process on the sphere, it is necessary and sufficient that $EY_{nm} = \mu \delta_{n0}\delta_{m0}$ for $\mu$ real and $\text{cov}(Y_{nm}, Y_{n'm'}) \stackrel{\text{def}}{=} E\{(Y_{nm} - EY_{nm})(Y_{n'm'} - EY_{n'm'})^*\} = c(n)\delta_{nn'}\delta_{mm'}$ with all $c_n$'s nonnegative and $\sum_{n=0}^{\infty}(n+1)c(n) < \infty$ [32].

To obtain axial symmetry, the restrictions on the covariance structure of the $Y_{nm}$'s are weaker: $EY_{nm} = \delta_{m0}\mu_n$ with $\sum_{n=0}^{\infty}\mu_n^2 < \infty$ and $\text{cov}(Y_{nm}, Y_{n'm'}) = c_m(n, n')\delta_{mm'}$ under suitable restrictions on the complex-valued covariances $c_m(n, n')$. Specifically, for $0 \leq m \leq N$, write $C_m(N)$ for the complex-valued (real-valued when $m=0$) covariance matrix of $(Y_{mm}, \ldots, Y_{Nm})'$. Then for all integers $0 \leq m \leq N$, $C_m(N)$ must be positive semidefinite. Unlike the homogeneous case, it does not appear possible to give simple necessary and sufficient conditions that guarantee convergence to a mean square continuous limit in (3). However, since we will only use models here for which $c_m(n, n') = 0$ whenever $\max(n, n')$ is greater than some fixed integer $N$, the mean square convergence of (3) will not be an issue.

The axially symmetric covariance function corresponding to (3) is given by

$$(4) \quad K(L, L', \ell) = \sum_{m=-\infty}^{\infty} \sum_{n,n'=|m|}^{\infty} e^{im\ell} \bar{P}_n^m(\sin L) \bar{P}_{n'}^m(\sin L') c_m(n, n'),$$

where $c_{-m}(n, n') = c_m(n, n')^*$. A desirable feature of using (4) to modeling axially symmetric processes is that, as [13] notes, *all* continuous axially symmetric covariance functions can be represented in this form, which follows from the completeness of $\bar{P}_m^m(\sin L), \bar{P}_{m+1}^m(\sin L), \ldots$ in the interval $[-\frac{1}{2}\pi, \frac{1}{2}\pi]$ for every $m$.

To gain some intuition into the interpretation of the $c_m(n, n')$'s, it is worth considering what restrictions on the properties of $Z$ follow from various restrictions on the $c_m(n, n')$'s. As we have already noted, if $c_m(n, n') = c(n)\delta_{nn'}$ for all possible $m, n, n'$, then $Z$ is homogeneous, which Figure 2 shows is badly untenable for total column ozone. We can obtain a somewhat richer class of models by considering $c_m(n, n')$ of the form $c_m(n)\delta_{nn'}$. Models of this form can have different levels of variation at different latitudes, but they are longitudinally reversible and have a reflection symmetry about the equator, $K(L, L', \ell) = K(-L, -L', \ell)$, neither of which are supported by Figure 2. A still weaker assumption is to take the $c_m(n, n')$'s to be real. However, it easily follows from (4) that this restriction implies longitudinal reversibility, so even this weaker assumption is not tenable.

Thus, it is not apparent how we might simplify the structure of the $c_m(n, n')$'s and still provide an adequate description of the covariance structure for this process. Fitting the infinite sum (4) to data obviously requires some further restriction. Here, we will consider the simple choice of setting

$c_m(n, n') = 0$ for $\max(n, n') > N$ for some positive integer $N$, or

$$(5) \quad K(L, L', \ell) = \sum_{m=-N}^{N} \sum_{n,n'=|m|}^{N} e^{im\ell} \bar{P}_n^m(\sin L) \bar{P}_{n'}^m(\sin L') c_m(n, n').$$

Since, for $m > 0$, the complex-valued positive semidefinite matrix $C_m(N)$ is of dimension $(N - m + 1) \times (N - m + 1)$, it requires $(N - m + 1)^2$ independent real parameters to specify, which, together with the $\frac{1}{2}(N + 1)(N + 2)$ real parameters needed to specify $C_0(N)$, yields a total of $1^2 + \cdots + N^2 + \frac{1}{2}(N + 1)(N + 2) = \frac{1}{3}(N + 1)(N^2 + 2N + 3)$ real parameters. This rapid growth with $N$ makes it difficult to take $N$ even moderately large and here we just consider $N = 6$ and 7.

To enforce the restriction that $C_m(N)$ is positive definite for each possible $m$, we will parameterize $C_m(N)$ using its Cholesky decomposition. For $m = 0$, this gives the familiar decomposition of the form $AA^T$, where $A$ is a real lower triangular matrix with nonnegative diagonal entries, and for $m > 0$, one gets a similar decomposition of the form $AA^*$ (the $^*$ indicates conjugate transpose), where $A$ is complex and lower triangular with (real) nonnegative diagonal entries. As in the real case, every positive semidefinite matrix has such a decomposition and the decomposition is unique if $C_m(N)$ is positive definite ([10], page 114). Denote by $A_m$ the lower triangular part of the Cholesky decomposition of $C_m(N)$, keeping in mind that $A_0$ is real-valued. Finally, we will include a nugget effect in our model, for a total of 120 parameters for the covariance function when $N = 6$ and 177 when $N = 7$.

**5. Data analysis.** To demonstrate that the model (5) with $N = 7$ can produce a fitted covariance function that matches many of the features of the empirical variograms in Figure 2, we will minimize a weighted least squares criterion to estimate the parameters of the model, despite my reservations about estimating covariance functions based on fits to empirical variograms ([25], Section 6.9). Although weighted least squares avoids any matrix inversions, if one uses fairly tight bins, the computations are still formidable due to the fact that here the variogram depends on two latitudes and a difference of longitudes. To reduce computations, not all pairs of points within an orbit were used; the first point in each pair was restricted to be in the latitude range $[10p, 10p + 1]$ for $p = -7, -6, \ldots, 8$ and the second point was restricted to be within 9° latitude and 20° longitude of the first. Details of the binning, the weights and the computations are given in the Appendix.

It is not possible to estimate all 177 parameters of the covariance function from the variogram. Specifically, since $\bar{P}_0^0$ is the constant function, changing $\operatorname{var} Y_{00}$ and $\operatorname{cov}(Y_{00}, Y_{0n})$ for $n = 1, \ldots, 7$ changes $K(L, L', \ell)$ by a function of the form $a(L) + a(L')$, which, as noted in the previous section, means the corresponding variogram $\gamma(L, L', \ell)$ is unaffected. Thus, we will only be estimating 169 parameters in this model, not 177.

The R function nlm was used to minimize the weighted sum of squares criterion as a function of the $A_m$'s and the nugget variance. It turns out that allowing the diagonal elements of the $C_m(N)$'s to be negative so that there are no constraints on the parameters speeds convergence considerably, although at the cost of a trivial lack of identifiability in the parameters [any column of any $C_m(N)$ can be multiplied by $-1$ without changing the model]. Even with this improvement, convergence is still very slow and this algorithm cannot be used routinely for a problem of this type. There is a substantial literature on least squares problems subject to a positive-semidefinite constraint [11, 15, 31] and it is likely possible to do better, but I will not pursue this issue further. In terms of fitting the qualitative features of the empirical variogram, the estimated model does fairly well. Figure 2 shows the contours for the fitted model and the empirical variogram at selected values for the first latitude. In many respects, the fit is quite good, capturing the lower levels of variation near the equator, the asymmetries between the Northern and Southern hemispheres and the differing levels of departure from longitudinal reversibility at different latitudes. Clearly, there is some misfit as well. In particular, at all latitudes, the very local variation is overestimated, a problem which should be resolvable by putting greater weight on shorter lags in the criterion function. The model also has trouble capturing the extended contours as one heads southward when $L = 40°\mathrm{S}$. All in all, though, the agreement is quite remarkable given the modest value for $N$. Furthermore, when $N = 6$, the fit is almost as good with an increase in the weighted least squares criterion of under 4%.

Despite this good agreement with some of the global features of the variogram, there are some serious problems with this fitted model that are not readily apparent in Figure 2. In particular, 28 out of 36 of the diagonal entries of the estimated $A_m$'s are effectively 0 in the sense that setting all 28 of these to exactly 0 increases the weighted least squares criterion function by less than 1 part in a million. Defining $V_{jm} = E(|Y_{jm} - EY_{jm}|^2 \mid Y_{mm}, \ldots, Y_{j-1,m})$, a zero in the $(j - m + 1)$th diagonal element of $A_m$ corresponds to $V_{jm} = 0$. Denoting by $a_m(i, j)$ the $(i, j)$th element of $A_m$, the nonzero diagonal elements of the $A_m$'s are $a_0(2, 2), a_0(3, 3), a_0(4, 4), a_1(1, 1), a_2(1, 1), a_2(2, 2), a_3(1, 1)$ and $a_4(1, 1)$, which correspond to positive values for $V_{10}, V_{20}, V_{30}, V_{11}, V_{22}, V_{32}, V_{33}$ and $V_{44}$. Thus, $Y_{jm}$'s with $m = 5, 6, 7$ do not contribute at all to the fitted model and for $m = 1, 3, 4$, $C_m(N)$ has rank one, which is in stark contrast to what happens for homogeneous models, in which case $C_m(N)$ is proportional to the identity matrix. It turns out that if $a_m(i, i) = 0$, then there is no loss in generality in the resulting class of covariance matrices in taking $a_m(j, i) = 0$ for $j > i$. The actual fitted model then arguably has only 65 "active" parameters: the nugget, 8 nonzero $V_{jm}$'s, the real values of $a_0(j,k)$ for $k=2,3,4$ and $j=k+1,\ldots,7$ and the real and imaginary parts of $a_1(j,1)$ for $j=2,\ldots,7$, $a_2(j,1)$ for $j=2,\ldots,6$, $a_2(j,2)$ for $j=3,\ldots,6$, $a_3(j,1)$ for $j=2,\ldots,5$ and $a_4(j,1)$ for $j=2,3,4$. Furthermore, setting $a_2(2,2)=0$ increases the criterion function by only 0.002%, so one can remove another 9 parameters with hardly any impact on the least squares fit. For $N=6$, the resulting fit only has 25 active parameters and, as already noted, this fit is only modestly worse than for $N=7$. Thus, the series expansion approach provides a quite parsimonious description of the larger scale features of the empirical variograms. However, the fitted model for $Z(L,\ell,t)$ (at a given $t$) is a nugget effect plus a function of rank 13 (i.e., the continuous part of $Z$ is determined by the real-valued $Y_{10}, Y_{20}$ and $Y_{30}$ and the complex-valued $Y_{11}, Y_{22}, Y_{32}, Y_{33}$ and $Y_{44}$), which is highly implausible. Thus, despite the respectable fit to the empirical variogram in Figure 2, the resulting fitted model is in some regards seriously wrong. We will explore this issue further in Section 6.

If one does not require $C_m(N)$ to be positive semidefinite, then because $\gamma$ is linear in the elements of the $C_m(N)$, the weighted least squares problem is linear and, hence, trivially solvable in closed form. For the present model with $N=7$, removing the positive definite constraint allows one to find estimates that reduce the weighted sum of squares by 68%, but, of course, the resulting model is not positive definite and is, in fact, "badly" so in the sense that the $C_m(N)$'s have many large negative eigenvalues. Thus, it is not surprising that the weighted least squares solution with the positive definite constraint should have some parameter estimates on the boundary of the parameter space. That so many of the parameter estimates should end up on the boundary was unexpected.

**6. Likelihood fits.** To carry out a likelihood analysis on the full dataset under a Gaussian model, one would need to model the space–time covariance structure of all of the observations. We can avoid modeling the temporal behavior by just considering the first orbit and acting as if all observations in that orbit were taken simultaneously. Even so, the number of observations, $s$, in this orbit is 13,216, so that a brute force calculation of the likelihood function at a given parameter value would require an $O(s^3)$ calculation and $O(s^2)$ memory, which would be very taxing computationally. However, for the model (5), it is possible to calculate the likelihood function at a given parameter value exactly using only a matrix decomposition on a (real-valued) matrix of rank $(N+1)^2$, as well as some matrix operations requiring $O(s(N+1)^4)$ operations that can be done once independent of the parameter values. The point is that our (real-valued) covariance matrices are of the form of a multiple of the identity matrix (from the nugget effect) plus a matrix of rank at most $(N+1)^2$, independent of the number of observations, which follows from the series representation for $K$ in (4). Cressie and Johanesson [4] exploited this fact to calculate kriging predictors based on the roughly 170,000 total column ozone measurements available in a day (in effect, ignoring the differences in time for observations from different orbits). Using the Sherman–Morrison–Woodbury identity [9] and a similar result for the determinant of matrices with this structure, we can then calculate the inverse and determinant of the covariance matrix needed for the Gaussian likelihood from the Cholesky decomposition of an $(N+1)^2 \times (N+1)^2$ matrix, requiring roughly $\frac{1}{3}(N+1)^6$ floating point operations.

To be more specific, we will assume the residual ozone process is a mean 0 Gaussian process. To evaluate the likelihood of the fitted model from the previous section, we need to specify values for $a_0(j,1)$ for $j = 1, \ldots, 8$ and we set them equal to 0 for simplicity. If we instead fit a pure nugget effect to the observations (i.e., treat them as Gaussian white noise) and maximize the likelihood with respect to the nugget variance, the loglikelihood is increased by over 21,000, or over 1.6 loglikelihood units per observation greater. Of course, since these data clearly show spatial dependence, the white noise model is itself terrible, so the weighted least squares fit is a truly awful description of the data, at least in terms of likelihood. Although part of the problem with the weighted least squares estimates may be due to setting $a_0(j,1) = 0$ for $j = 1, \ldots, 8$, the white noise model has no parameters for spatial dependence, so it cannot be the whole problem. Thus, we have a truly stunning example of the potential problems of using agreement with empirical variograms as a way of fitting spatial covariance functions.

The estimated nugget variance is $5.36 \times 10^{-3}$ under the white noise model, an order of magnitude larger than the estimated nugget under the weighted least squares criterion. As noted in the previous section, even the value $5.33 \times 10^{-4}$ from the weighted least squares fit appears to be too large based on what is shown in the empirical variograms, which suggest a nugget variance of no more than $1.7 \times 10^{-4}$ at all nonpolar latitudes (for latitudes between 50°N and 50°S, the smallest binned variogram value is between $1.2 \times 10^{-4}$ and $1.7 \times 10^{-4}$). And, indeed, a model that provided a good description of the local variation of the data should yield an estimated nugget of around this size. The white noise model ignores the spatial dependence and uses the empirical variance (uncentered, since we have assumed the process has mean 0) of all of the observations to estimate its lone parameter. If we attempt to fit the model from the previous section to the data from the first orbit via maximum likelihood, we get an increase in loglikelihood of 7372 units over the white noise model. This model has a nugget of $1.69 \times 10^{-3}$, which is still radically higher than what the empirical variograms show. This huge discrepancy between the MLE (maximum likelihood estimate) and the empirical variogram can only be explained by the inappropriateness of the chosen model. The fact that the MLE of the nugget is far too large indicates the source of the problem is that the very smooth spherical harmonics of degree and order at most 7 cannot accurately describe the smaller-scale dependencies of the process. How much larger $N$ would need to be to provide a decent description of the data in terms of likelihood is unclear, but as the number of operations needed to do the necessary matrix calculations for a single likelihood evaluation grows like $N^6$ and the number of parameters like $N^3$, we cannot take $N$ all that large before computations requiring optimizing or integrating over the exact likelihood become overwhelming.

To get a better idea of how the truncated series expansion model compares to a more sensible model than white noise, it is helpful to look at a small enough subset of the data so that exact likelihood calculations can be done for other Gaussian models. Specifically, let us consider the 839 observations taken during the first orbit on May 1, 1990 between the latitudes of 65°S and 55°S. The process looks reasonably isotropic in this latitude range (not shown, but similar to the variogram at 60°N), so, in addition to evaluating the likelihood of the previous weighted least squares fit and the MLE of the white noise model for these 839 observations, we also find the MLE under a three-parameter model for the covariance structure including a nugget variance and a term of the form $\theta_1 \exp(-d/\theta_2)$ with $d$ the chordal distance between observations. The maximized loglikelihood of the white noise model is 204 greater than the weighted least squares fit from the previous section with $N = 7$ and the loglikelihood of the nugget plus exponential model is 1403 greater than the white noise model. The maximized loglikelihood under (5) with $N = 7$ (and 177 parameters) is 364 less than under the exponential model. Many of the parameter estimates for (5) are unstable, as might be expected for a model meant for a global scale with 177 parameters when there are only 839 observations in a fairly small region. However, the estimated nugget does appear to be numerically stable and, while it is much smaller than the MLE under the white noise model (around $1.1 \times 10^{-3}$ rather than $1.3 \times 10^{-2}$), it is still much larger than the apparent nugget shown in the empirical variogram. In the nugget plus exponential model, the estimated nugget is $1.95 \times 10^{-4}$, which is only moderately larger than the empirical nugget effect.

**7. Discussion.** Given that total column ozone levels can change substantially on the time scale of a few hours, the loss of within day temporal information in the Level 3 TOMS could be important in some applications, for example, setting initial conditions in numerical models for ozone [17] and, as noted in the Introduction, for mapping surface ultraviolet radiation. Thus, it might be helpful if a gridded version of the TOMS data were available in which the times of observations were preserved. For grid cells covered by more than one orbit in a given day, there would then be more than one observation reported on that day, each with its own time. A plausible name for such a product would be Level 2.5 TOMS. Yet a further embellishment would be to make a multiple imputation version of such a data set, which would then allow users to account for the uncertainty in these interpolated values. Such multiply imputed products have in fact been called for recently in the meteorological literature [24].

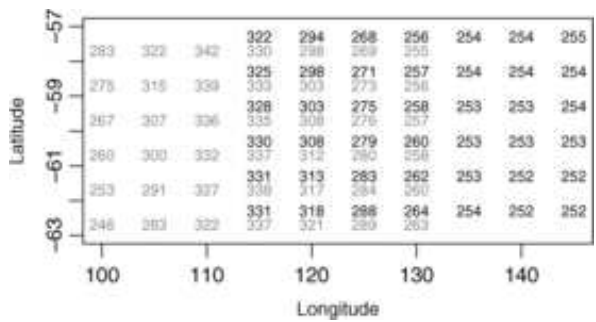

FIG. 5. *Predicted ozone levels (Dobson units) on grid May 1, 1990. Predictions for third orbit in black, for fourth in gray, latitudes for two orbits slightly offset to avoid overlaps.*

Figure 5 displays a simple example of such a gridded data set for May 1, 1990 for latitudes between 62.5°S and 57.5°S based on the third and fourth orbits on that day (the second orbit had missing observations). For example, the black numbers in Figure 5 were obtained as follows: using observations from the third orbit in latitude range 65°S to 55°S and the covariance function for the exponential model with parameter estimates from the previous section (i.e., estimated from the first orbit on that day), kriged values of the residual field were computed on a 1° latitude by 5° longitude grid, the mean field added back and the result exponentiated to obtain ozone values for a range of longitudes covered by the orbit. These predicted values should be considered predicted medians rather than means, for which one would want to take into account the nonlinear transformation (see [3], page 135). The gray numbers in Figure 5 were obtained similarly using the fourth orbit. We see that, in this case, there are modest differences in the predictions for the two orbits where they overlap, but in other cases, the differences can be substantially larger.

The computational advantages of representing the covariance structure as a diagonal matrix plus a matrix of fixed rank (i.e., independent of the size of the data set) are very substantial, both for computing kriging predictors and for calculating likelihoods. Thus, it is worth considering how this approach might be modified to retain at least some of the computational gains but at the same time provide a better representation of the small-scale behavior of the total column ozone process. Here are three possibilities that could be worth exploring. First, increase the value of $N$, the maximum value for $n$, but do not include every $m$ such that $|m| \leq n$ in (3), so that some higher frequency terms can be included without increasing the rank of the continuous part of the process too dramatically. Second, replace the basis functions by less smooth functions, which is what Cressie and Johanneson [4] do. However, their models are not axially symmetric and, in any finite expansion, replacing the sines and cosines with other functions of longitude loses the axial symmetry. Thus, using less smooth functions of longitude would require sacrificing exact axial symmetry. Replacing the Legendre polynomials by, say, a wavelet expansion, may help to get more realistic high frequency variations across latitudes, though. A third possibility is to recognize that we can replace the diagonal part of the covariance matrix by any matrix $M$ for which linear systems $M\mathbf{x} = \mathbf{y}$ can be solved quickly and still gain a computational advantage using the Sherman–Morrison–Woodbury identity. Thus, we might replace the nugget effect by a covariance function that is identically 0 for points more than a modest distance apart, yielding covariance matrices for which linear systems can be solved quickly using sparse matrix methods [7].

Another possible solution to the computational problems posed by large spatial data sets is to abandon calculating likelihoods and kriging predictors exactly, using, for example, approximate likelihood methods described in [2, 6] and [27]. Of course, even if the computational problems can be handled, we are still faced with the challenge of finding a model that captures both the large and small scale features of the process. Thus, one could take the further step of abandoning the search for a single global model and develop separate models for different latitude bands, which would simplify both the modeling and computational challenges. However, the large-scale features may be of particular scientific interest, for example, one may want to predict the time evolution of the $Y_{nm}$'s for relatively small values of $n$ and $m$.

Modeling covariance functions using truncated series expansions can be viewed as an example of the "subset of regressors" approach to reducing the computation in nonparametric function estimation [22]. Although [30] and [22] have noted problems with this approach for functions with fine features, the analysis here perhaps highlights how wrong things can go using such models, especially in terms of their likelihoods.

Given how hard it is just to model the purely spatial variation in total column ozone residuals, what are the prospects for developing space–time models on a global spatial scale and daily time scale? As noted by Jones [13], the model (3) can be readily extended to the space–time setting by letting each $Y_{nm}$ be a stochastic process in time. If one was only interested in the large-scale features of this process, then such a model with only a limited number of terms might be fairly useful. However, if we wish to describe small-scale spatial features accurately, we may want to abandon the series approach and try something along the lines of [14]. As already noted, the Level 3 gridded data are not sufficient for distinguishing small-scale spatial and temporal variations. However, direct use of the Level 2 data in a comprehensive space–time model would certainly lead to tremendous computational challenges, so that a Level 2.5 data set might be a better starting point for such a study.

## APPENDIX

This appendix describes some of the computational details behind the work in this paper. To compute the Legendre polynomials $P_n^m$ for $0 \leq |m| \leq n \leq 7$, the method described in Section 4.4.4 of [33] was used to calculate and store $P_n^m(\sin\theta)$ for every $\theta$ between $-90°$ and $90°$ by increments of $0.25°$. Results for intermediate angles were obtained and stored using a cubic spline interpolator through these exact values.

To obtain the binned variogram at a given nominal latitude $L_0$, the following procedure was used (all angles are measured in degrees here). For integer pairs $(j,k)$ with $-9 \leq j < 9$ and $-20 \leq k < 20$, let $(L_i, \ell_i)$ and $(L_i', \ell_i'), i = 1, \ldots, q_{jk}$, be the pairs of observations in a common orbit satisfying $L_0 \leq L_i < L_0 + 1$, $j \leq L_i - L_i' < j + 1$ and $k \leq \ell_i - \ell_i' < k + 1$. Define $\bar{L} = q_{jk}^{-1} \sum_{i=1}^{q_{jk}} (L_i - L_i')$, $\bar{\ell} = q_{jk}^{-1} \sum_{i=1}^{q_{jk}} (\ell_i - \ell_i')$ and

$$(6) \qquad \hat{\gamma}\left(L_0 + \frac{1}{2}, L_0 + \frac{1}{2} + \bar{L}, \bar{\ell}\right) = \frac{1}{2q_{jk}} \sum_{i=1}^{q_{jk}} \{Z(L_i, \ell_i) - Z(L_i', \ell_i')\}^2.$$

Note that we have not assigned this average to the "center" of the bin, that is, $(L_0 + \frac{1}{2}, L_0 + j + 1, k + \frac{1}{2})$. Given the nature of the observation pattern for TOMS, there are sometimes substantial differences between $(L_0 + \frac{1}{2}, L_0 + \frac{1}{2} + \bar{L}, \bar{\ell})$ and the bin center; the right-hand side of (6) should generally be more nearly unbiased for $\gamma(L_0 + \frac{1}{2}, L_0 + \frac{1}{2} + \bar{L}, \bar{\ell})$ than for $\gamma(L_0 + \frac{1}{2}, L_0 + j + 1, k + \frac{1}{2})$.

The weights in the weighted least squares procedure were set to the number of pairs of observations contributing to each bin divided by the angle between the bin centers plus $1°$. This weighting is obviously somewhat arbitrary; dividing by angle plus $1°$ gives more weight to shorter arc distances, although as noted in Section 5, perhaps even greater weight should have been given to shorter lags.

DEPARTMENT OF STATISTICS
UNIVERSITY OF CHICAGO
CHICAGO, ILLINOIS 60637
USA
E-MAIL: stein@galton.uchicago.edu